\documentstyle[prl,epsf,floats,aps,amsfonts,twocolumn]{revtex}

\newcommand {\be}{\begin{equation}}
\newcommand {\ee} {\end{equation}}
\newcommand {\ba}{\begin{eqnarray}}
\newcommand {\ea} {\end{eqnarray}}

\def \F2 {FPL${}^2$ }

\def \OMIT #1{}
\def \rem #1 {{\it #1}}

\begin{document}

\twocolumn[\hsize\textwidth\columnwidth\hsize\csname
@twocolumnfalse\endcsname

\title {The traveling salesman problem, conformal invariance, and dense polymers}
\author {J.L. Jacobsen,$\!^1$ N. Read,$\!^2$ and H. Saleur$\,^{3,4}$}

\address{$^1$ Laboratoire de Physique Th\'eorique et Mod\`eles Statistiques,
Universit\'e Paris-Sud, B\^atiment 100, F-91405 Orsay,
France\\$^2$ Department of Physics, Yale University, P.O. Box
208120, New Haven, CT 06520-8120, USA\\$^3$ Service de Physique
Th\'eorique, CEA Saclay, 91191 Gif sur Yvette, France\\
$^4$ Department of Physics and Astronomy, University of Southern
California, Los Angeles, CA 90089-0484, USA}

\date{May 29, 2004}
\maketitle

\begin{abstract}
We propose that the statistics of the optimal tour in the planar
random Euclidean traveling salesman problem is conformally
invariant on large scales. This is exhibited in power-law behavior
of the probabilities for the tour to zigzag repeatedly between two
regions, and in subleading corrections to the length of the tour.
The universality class should be the same as for dense polymers
and minimal spanning trees. The conjectures for the length of the
tour on a cylinder are tested numerically.

\end{abstract}
]



The traveling salesman problem (TSP) is a classic problem in
combinatorial optimization. The basic problem is, given a set of
$N$ marked points (``cities'') in the plane, to find the closed
path (a cycle or ``tour'') of shortest length that passes through
each city once. In the random uniform TSP, the cities are chosen
randomly, and are independently and uniformly distributed over
some bounded domain $\cal D$, say a square, with mean density
$\bar{n}$.
While much effort has been expended on finding algorithms that
produce the optimal tour for a given set of cities, for
statistical physicists the interest of the problem is in the
statistical properties of the optimal tour in the random uniform
problem, to which we will refer simply as TSP
\cite{review,Steele}. In the past, much attention has been given
to the total length $\ell$ of the optimal tour, which for $N$
cities in a square behaves as $\ell(N) \sim \beta A/a$ with
probability one as $N\to\infty$ \cite{Hammer,twiddle}, where $A$
is the area of the square, $a=\sqrt{A/N}\equiv1/\sqrt{\bar{n}}$ is
the typical spacing of the cities, and $\beta$ is a constant:
$\beta \simeq 0.7120$ \cite{PerMar}. Finite $N$ corrections to the
mean length in a cube of dimension $d$ with periodic boundary
conditions have been studied \cite{PerMar}.

In this paper, we consider geometrical properties of the optimal
tour, other than the mean length for the square. These properties
include the dependence of the mean length on the aspect ratio when
the cities are distributed in a rectangle or on the surface of a
cylinder. They also include the connectivity of the path, which we
quantify by defining the number of times the tour alternately
enters (or ``zigzags'' between) two specified subregions. We
formulate conjectures based on statistical conformal invariance
\cite{cft} of the properties of the optimal tour over length
scales much larger than $a$. For clarity we separate these
conjectures and call them TSPI, II, and III. The conjectures are: I)
{\em conformal invariance} of the distribution of tours, and hence
power-law behavior of the probability $P_k(r)$ for zigzagging $k$
times between two regions that are a distance $r$ apart (and
much further from the boundary of ${\cal D}$),%
\be%
P_k(r)\propto r^{-2x_k} \label{pkr}
\ee%
for $r\gg a$,  for some exponents $x_k$. This implies that the
optimal tour is a random fractal, and the $x_k$ determine the
fractal dimensions $D_k=2-x_k$ of various sets associated with the
tour. These predictions, including the values of $x_k$, are {\em
universal}; they do not depend on the precise distribution of
cities, which might even be correlated, as long as the
distribution is translationally and rotationally invariant, and
any correlations are short-range; II) the mean length
$\overline{\ell}$
of the optimal tour in a domain of area $A$ that has a smooth
boundary of length $C$ has the form
$\overline{\ell}/a=\beta\bar{n}A+\gamma C/a+\ldots$ (a weaker
version of this has been proved for the square \cite{Rhee}, and
implies that $\gamma>0$). If we define $\Delta \ell/a\equiv
\ell/a-\beta\bar{n}A-\gamma C/a$, then we expect %
\be%
\overline{\Delta\ell}/a \sim -\frac{1}{6}\lambda c\chi\ln(|{\cal D}|/a) \label{f}
\ee%
as $N\to\infty$, where $\lambda>0$ and $c$  are constants, $|{\cal D}|$
is the diameter of $\cal D$, and $\chi$ is the Euler number
of $\cal D$ ($\chi=2-2h-b$, where $h$ is the number of handles, $b$ is
the number of boundaries). If the tour is also constrained to zigzag $k$ times
between two fixed regions far from the boundary, then we expect in addition%
\be%
\overline{\Delta\ell}/a\sim 2\lambda x_k \ln (r/a)%
\label{ell}
\ee%
for $r/a\to\infty$  [here $x_k$ are the same as in eq.\
(\ref{pkr})]. The values of $\beta$, $\gamma$, and $\lambda$ are
not universal; III) the universality class for the TSP is the same
as that known as dense polymers, so the exact values are \cite{Nienhuis82,Saleur86,DupSal87,Batchelor88}%
\be %
x_k=(k^2-4)/16, \quad c=-2.%
\label{xk}
\ee %
After explaining these conjectures further, we study
the length as in TSPII numerically by a transfer-matrix-like
approach on a cylinder, testing the conformal symmetry proposed in
I, and finding reasonable agreement with the quantitative
conjectures in II and III.

There is a possible relation with the minimal spanning tree (MST)
problem (given $N$ cities, find the tree of smallest length with
those cities as vertices; the cities are chosen at random as in
the TSP).
Analogs of parts of our TSPI and III have already been discussed
for MSTs \cite{ABNW,MST}. Here the analog of TSPIII would involve
so-called uniform spanning trees (USTs) in place of dense
polymers. The equivalence of USTs and MSTs in $d=2$ is not
excluded by rigorous results \cite{ABNW}, and in our view is {\em
supported} by existing numerics \cite{MST}. A tree in two
dimensions is equivalent to a nonintersecting loop (take the
boundary of a ``thickened'' tree), and the universality classes of
USTs and dense polymers are the same in $d=2$ \cite{dup} (and also
the same as stochastic Loewner evolution at parameter value
$\kappa=8$ \cite{schramm}). Hence in two dimensions we expect the
universality classes for TSP and MSTs to be the same (the
Hausdorff dimension studied in Ref.\ \cite{MST}, which is equal to
5/4 for USTs \cite{Maj}, corresponds to our $D_4$).

We now further explain our predictions. First, we note that the
optimal tour cannot cross itself, as that
would allow a shorter tour to be found. For a typical set of
cities, the tour comes close to every point in the region
considered. In the scaling limit $a\to0$ for a fixed $A$, the path
becomes a random space-filling (Peano) curve with Hausdorff
dimension $D=D_2=2$ \cite{Hammer}. For a tour within a
simply-connected domain, the interior of the curve is well
defined, and can be shaded black, leaving the exterior white. The
interior and exterior form interlocking trees which appear
statistically alike, except near the boundary of the domain. This
implies a self-duality to the tour. One strongly suspects that
such a random self-dual curve must be scale, and very likely also
conformally, invariant, in a sense we must now explain.

Consider a square at arbitrary position well within the interior,
with side $L$ much less than the size of the domain, and with
edges parallel to the coordinate axes. The tour passes through
this square some number of times, entering and leaving on two
sides (not necessarily distinct) of the square. We examine the
number $n_x\geq 0$ of times the tour {\em crosses} the square in
the $x$-direction, that is the number of segments of the tour that
have one end on each of the two edges parallel to the $y$-axis.
Similarly, we can consider the number of times $n_y$ it crosses
the square in the $y$ direction. If $n_x>0$, then $n_y=0$, and
{\it vice versa}. We expect that the joint probability
distribution for $n_x$, $n_y$ is concentrated at small values of
$n_x$, $n_y$. Then the expectation $\overline{n_x}$ will be of
order 1. As $L/a$ increases, $\overline{n_x}$ will remain nonzero,
as the tour must occasionally cross the square. (The possibility
that the probability of $n_x=n_y=0$ approaches 1 appears to be
excluded, because of the requirement that the tour be a single
cycle.) Thus we expect that the joint probability distribution for
$n_x$ and $n_y$ is scale invariant for large $L/a$.

Consider two disks $\cal A$, $\cal B$ of radius $r_0$, the centers
of which are separated by $r>2r_0$. Let $P_k(r)$ be the
probability for crossing (zigzagging) between the two disks
precisely $k$ times ($k$ even); more precisely, it is the
probability that $k$ distinct connected segments of the tour lying
outside both disks have one end on the boundary of each disk.
By standard arguments, scale invariance of the
crossing probabilities (applied to annuli concentric with $\cal A$
or $\cal B$) leads us to expect eq.\ (\ref{pkr}) to hold as a function
of $r$, for $r\gg a$ and $a$, $\cal A$, $\cal B$, fixed (and with $r$
much less than the distance of $\cal A$ or $\cal B$ from the
boundary of $\cal D$), and that $x_k>0$ for $k>2$. With $D_2=2$,
it follows that $x_2=0$.


We may define similar crossing or ``$k$-leg'' probabilities for
$k$ odd by allowing the path to end at any two of the marked
cities, such that the total length is minimized. For open paths,
we define $P_k(r)$ for $k$ odd as the probability that the path
starts in $\cal A$ and ends in $\cal B$, crossing between them $k$
times. We expect $x_1<0$, meaning that the optimal path will
usually have its ends far apart. These definitions of the 2-point
correlations can be easily extended to general $n$-point functions
which are probabilities for the path to pass between $n$ disks in
some specified sequence. As for $n=2$, we expect these to possess
scaling limits as $a\to0$, and these define a probability
distribution on non-self-intersecting self-dual space-filling
curves, which will be universal, and which we wish to
characterize. The definition of the $k$-leg events can be
generalized to the case when a disk is close to a boundary of the
domain, which is assumed here to be straight. For $n=2$, if the
distances of $\cal A$ and $\cal B$ to the boundary are both much
less than their separation $r$, then we expect $P_k(r)\propto
(r/a)^{-2\widetilde{x}_k}$, with universal exponents $\widetilde{x}_k$
different from $x_k$ (again, $\widetilde{x}_1<0$, and
$\widetilde{x}_2=0$).


Conformal invariance is expected in scale-invariant statistical
problems when they are defined by {\em local} processes. In the
TSP, the length which is to be minimized is local in the sense
that it is the sum of small local steps. There may be a concern
that the global constraint of visiting each city once violates
locality. However, such a condition is also present in dense
polymers, so is not necessarily an issue.

For the following arguments, and for numerical purposes, it is
convenient to consider the TSP on a cylinder, with circumference
$W$ (i.e.\ in the region $0<x<W$ in the plane with a periodic
boundary condition in the $x$-direction), and length $L$, so
$A=LW$. The cities are uniformly distributed over this region. In
this case, scaling arguments suggest that the probability that the
path crosses at least $k$ times between two regions within $r_0$
of the ends behaves for $N\to \infty$ as $P_k(W,L)\propto e^{-2\pi
x_k L/W}$ for $L/W\gg 1$. If conformal invariance holds, then the
exponents $x_k$ here are the same as those defined above in the
plane, by using a conformal mapping of the plane to the cylinder
\cite{gaps}.

Now suppose that, instead of the tour being unconstrained, the
tour is {\em required} to cross at least $k$ times between the end
regions, no matter what the positions of the cities. As the tour
must minimize its length, for $k>2$ the mean length
$\overline{\ell_k}$ of the constrained tour will be greater than
or equal to that of the unconstrained ($k=2$) one, by an amount
$\propto L$. In fact, from scale invariance we expect that, for
each $k$, as $N\to\infty$ with $L/W$ fixed,
$\overline{\Delta\ell_k}/a\equiv\overline{\ell_k}/a-\beta\bar{n}
A-2\gamma W/a$ is proportional to a universal function of $L/W$,
and this function is $\propto L/W$ as $L/W\to\infty$. We further
expect that, for $k$ even, the change in length of the constrained
tour $(\overline{\ell_k}-\overline{\ell_2})/a$ is proportional to
the logarithm of the probability for $k$ crossings in the
unconstrained case: $(\overline{\ell_k}-\overline{\ell_2})/a \sim
-\lambda\ln P_k(W,L)$ as $a\to 0$. Thus for large $L/W$, we have%
\be%
(\overline{\ell_k}-\overline{\ell_2})/a\sim 2\pi\lambda x_k L/W,%
\label{x}
\ee%
where $\lambda\geq 0$ is a non-universal constant, but is the same
for all $k$, and we expect this form to hold for $k$ odd also [conformally
mapping this to the plane yields eq.\ (\ref{ell})]. We
expect similar behavior also for the unconstrained tour, and so we
define a universal constant $c$ by
\be%
\overline{\Delta\ell_2}/a\sim -\frac{\pi\lambda c}{6} L/W.%
\label{c}
\ee%
One expects then $\lambda>0$ and $c\leq 0$.

The appearance here of only a single non-universal constant
$\lambda$, and the various scaling forms suggested, require
further explanation. We are using an analogy with conformal field
theories (CFTs) in two-dimensional critical phenomena. There, the
variation of the free energy (the logarithm of the partition
function) with respect to the metric of the space defines the
stress tensor of the theory. This leads to universal scaling forms
for the subtracted free energy in various geometries or with
constraints or sources imposed, as in a correlation function
\cite{gaps,central,CarPes}. Our central conjecture is that, up to a
non-universal factor $\lambda$, the length behaves as the free
energy in some CFT ($c$ is then the central charge). The reason is
that the length $\ell/a$ is the integral over space of a local
density, and in many CFTs (including the dense polymer theory
considered below) the stress tensor is (up to a factor) the only
local spin-2 operator of the correct symmetry and lowest scaling
dimension that can appear as the variation of the length with
metric. As free energy differences also determine probabilities
for events, the relation of the length differences with the
probability also follows. By such arguments in the plane, or in
other smooth domains (including non-planar ones, such as the sphere or
torus), one obtains the scaling form (\ref{f}) \cite{CarPes}.
For geometries in which $\chi$ is zero, the term in eq.\ (\ref{f}) is replaced
by $\lambda$ times a universal scale-invariant functional of the
geometry \cite{central}, as for the cylinder in eq.\ (\ref{c}).


There is a probability distribution for non-self-intersecting
self-dual space-filling curves that arises from statistical
mechanics. This is the Nienhuis dense-polymer phase that
originally arose from the low-temperature phase of the O($m$) loop
model at $m\to0$ \cite{Nienhuis82}. It is a model of closed loops
on the honeycomb graph, on which each edge of the graph is
occupied at most once. The partition function of the model is
$Z=\sum K^E m^{\cal L}$, where the sum is over all such loop
configurations, $K$ plays the role of inverse temperature, $E$ is
the number of edges occupied, and $\cal L$ is the number of
distinct loops. When $m=0$, $Z=0$, but the partition function for
a single closed loop on the honeycomb graph can be obtained by
differentiating, $Z'=\partial Z/\partial m|_{m=0}$, and then
probabilities $P_k(r)$ can be defined analogously to those above.
The model has a critical point at $K=K_c=(2 + \sqrt{2})^{-1/2}$,
and the region $K_c<K<\infty$ is a conformally-invariant
low-temperature phase in which the scaling exponents are
independent of $K$. The scaling limit of the probability
distribution is highly robust: no local perturbations are
relevant, except for that of allowing the loops to
cross \cite{JRS}. Therefore we find it natural to propose TSPIII,
that the TSP and dense-polymer universality classes are the same.
For dense polymers, $c=-2$, and the $k$-leg exponents are given by
eq.\ (\ref{xk}) for the bulk, and by \cite{DupSal87}%
\be%
\widetilde{x}_k=k(k-2)/8%
\ee
for the boundary.

We have studied the length of a tour on a cylinder by extensive
numerical calculations. The approach is similar to transfer matrix
methods in statistical mechanics. The tour on a cylinder is
``grown'' starting from one end of the cylinder. The distance $y$
along the cylinder plays the role of time and will be denoted $t$
from here on. We set $a=1$. The transfer process starts with a
city $p_1$ at $x_1=0$, $t_1=0$ from which $k$ lines emerge (we say
it has valence $k$), and terminates with a similar city $p_N$ at
some $x_N$ and $t_N$. The remaining $N-2$ cities $p_i$ have
coordinates $x_i$, $t_i$, and valence $2$. The variables $x_i$ and
differences $t_i-t_{i-1}$ are all independent for $i>1$. The
$x_i$'s are chosen uniformly in $[0,W]$, while the differences
$t_i-t_{i-1}>0$ are exponentially distributed with mean $1/W$
(this reproduces the Poisson process with density 1).

For a given set of cities, and for each time $t_i$, the
information needed to complete a tour and find the optimal one
consistent with the constraint is encoded in a set of states, each
of which has a weight (length of path) associated with it. The
transfer matrix will evolve these states and weights to $t_{i+1}$.
A state consists of a list of the $M$ cities $p_{i_a}$, $a=1$,
\ldots, $M$ ($i_1<i_2<\cdots i_M\leq i$), whose valence has not
yet been satisfied (by connection to other cities), plus
connections among these cities. Of the $M$ cities, there may be
some that are not connected to any other one, some that are
connected to just one other, and there is always one distinguished
set of at most $k$ cities that are connected to each other. To
each of these states, we can associate the set of paths (with
minimum total length and no closed loops) that form the
connections and satisfy the valence of all the cities $p_j$,
$j\leq i$, not in the list; the distinguished set of $k$ cities
are connected to the initial city $p_1$ (which itself will be in
the list during the early stages). The length of this set of paths
is the weight associated with the state.

When a city $p_{i+1}$ of valence $2$ is added, the states and
their weights at $t_{i+1}$ are related to those at $t_i$ by one of
three moves: (1) $p_{i+1}$ may be unconnected to any other city;
(2) $p_{i+1}$ may be directly connected to one other city
$p_{i_a}$ in the list; (3) $p_{i+1}$ may be directly connected to
two other cities $p_{i_a}$, $p_{i_b}$, so it will not appear in
the list (if $p_{i_a}$, $p_{i_b}$ were connected to each other at
time $t_i$, this move is forbidden). In moves (2) and (3), cities
$p_{i_a}$ or $p_{i_b}$ whose valence becomes satisfied disappear
from the list, and the connections of other cities change
accordingly. The weight of a state at time $t_{i+1}$ is equal to
that of the state it came from at time $t_i$ plus the length of
the connections that have been added. If a state at time $t_{i+1}$
can arise from more than one state at time $t_i$, its weight is
taken as the minimum of the various possibilities. At the final
time $t_N$, a $k$-valent city is added (using similar moves), and
we take the length $\ell_k$ to be the length of the state at $t_N$
in which all valences are satisfied (the condition of not
connecting already-connected cities is dropped at this step).

Clearly, with the above moves, the size of the state space grows
without limit as $i\to\infty$. In the spirit of heuristic (local
optimization) algorithms we deal with this by discarding all
states at each time $t_i$ that contain a city $p_{i_a}$ with $i_a
< i-i_{\rm max}$. The transfer matrix is then finite dimensional,
with a size that grows exponentially with $i_{\rm max}$. This
truncation means that steps with a long $t$-component are
suppressed. For $i_{\rm max}\gg W$, they would be rare on the
optimal path anyway.

\begin{table}
 \begin{center}
 \begin{tabular}{l|lllll}
             &$W=2$ & $W=3$ & $W=4$ & $W=5$ & $W=6$ \\ \hline
  $\overline{\ell_2}/(LW)$   & 1.07497(2) & 0.8330(1) & 0.7571(1) & 0.7330(2) & 0.720(2)\\
  $\overline{\ell_1}/(LW)$   & 0.7457(4) & 0.7175(3) & 0.715(1)& 0.713(2)& \\
  $\overline{\ell_3}/(LW)$   & 1.5286(1) & 1.075(1) & 0.872(1) & 0.80(1) & 0.76(1) \\ \hline
  $\lambda c$     & $-2.39$ & $-3.33$
             & $-2.98(1)$ & $-2.04(2)$ & $-2.03(15)$ \\
  $\lambda x_1$   & $-0.15(1)$ & $-0.20(2)$ & $-0.15(2)$ & $-0.10(5)$ & \\
  $\lambda x_3$   & &  0.30(2) &   0.33(2) &   0.29(1) &  0.26(2) \\
 \end{tabular}
 \end{center}
 \vspace{-4mm}
 \protect\caption[2]{\label{tab} Length $\overline{\ell_k}$ of tour per unit area, for each
 width $W$, extrapolated to $i_{max}\to\infty$, with values of
 $\lambda c$, $\lambda x_k$ deduced from $\ell_k$ for pairs $W$, $W-1$. Data
 for $W=1$ are not shown.}
 \vspace{-5mm}
\end{table}

We have produced results for $\ell_k(i_{\rm max})$ with $1 \le W
\le 6$ and $4 \le i_{\rm max} \le 8$ or $9$. In each case, we find
the optimal tour for $N=10^5\cdot W$ cities, and then average over
10 independent samples. The systematic error due to the finite
value of $N$ is negligible as compared to the statistical error
(sample-to-sample fluctuations). To extrapolate to the $i_{\rm
max}\to\infty$ limit we use the Ansatz $\overline{\ell_k}(i_{\rm
max})/(LW) = \overline{\ell_k}(\infty)/(LW)+A e^{-B i_{\rm max}}$,
$A$ and $B$ being $k$-dependent constants, which matches the data
very well, especially when $i_{\rm max} > W$. In Table~\ref{tab}
we display the extrapolated data for $k=2$, $1$, $3$. Estimates
for $\lambda c$ ($\lambda x_k$, $k=1$, $3$) were based on eq.\
(\ref{c}) [eq.\ (\ref{x})], using $\overline{\ell_2}$
($\overline{\ell_k}-\overline{\ell_2}$) for pairs $W$, $W-1$
($\gamma$ was neglected). We obtained a value $\beta=0.7119(3)$,
in good agreement with Ref.\ \cite{PerMar}. Our final best
estimates are $\lambda c=-2.0(2)$, $\lambda x_1=-0.15(5)$, and
$\lambda x_3=0.28(4)$. We note that the values $\lambda=1$,
$c=-2$, $x_1=-3/16$, and $x_3=5/16$ lie within the error bars. At
present, we have no explanation for why $\lambda$ should be close
to one.

Finally, for TSP in a three-dimensional region of fixed thickness
$L_3$ in the $z$-direction and a domain $\cal D$ of area $A\gg
L_3^2$ in the $x$--$y$ plane, projecting the tour into the plane
produces a two-dimensional problem, but the projected tour will
occasionally cross itself. If our conjectures hold for $L_3=0$,
then by Ref.\ \cite{JRS} any small, positive $L_3$ causes a
crossover to a ``Goldstone phase''. It then seems very
likely that TSP in all $d>2$ is also described by the Goldstone
phases, which would mean that the segments of the tour in a box of
side $L$ have Hausdorff dimension 2, and behave as Brownian walks
on large scales. Conformal invariance would be lost in these
cases.

We thank P.W. Jones, D.L. Stein, and A.A. Middleton for useful
discussions. This work was supported by NSF grant no.\
DMR-02-42949 (NR) and by the DOE (HS).




\begin{references}
\bibitem{twiddle}     Throughout this paper, we use notation $X\sim Y$ as
                       $Z\to\infty$ in the strict sense: $\lim_{Z\to \infty}X/Y = 1$.
\end{references}
\end{document}